\documentclass[nofootinbib,superscriptaddress,twocolumn]{revtex4}
\usepackage[percent]{overpic}

\usepackage{natbib}
\usepackage{graphicx}
\usepackage{epsfig}
\usepackage{amsmath}
\input{epsf}
\usepackage{psfrag}

\usepackage[usenames,dvipsnames]{color}

\newcommand{\beq}{\begin{eqnarray}}
\newcommand{\eeq}{\end{eqnarray}}
\newcommand{\Slash}[1]{{\ooalign{\hfil/\hfil\crcr$#1$}}}

\newcommand{\ket}[1]{\left|#1\right>}

\newcommand{\nn}{\nonumber \\}

\usepackage{amsmath,color}
\usepackage{amssymb}
\usepackage{bm}
\usepackage{graphicx}
\usepackage{multirow}

\begin{document}
\begin{flushright}
{\small
TUM-HEP-1611/26
}
\end{flushright}

\title{Three-qubit entanglement in the Bethe-Heitler process
}

\author{Haotian Cao}
\affiliation{Department of Physics \& Astronomy,
Northwestern University, Evanston, Illinois 60208, USA} 
\affiliation{Center for Frontiers in Nuclear Science, Stony Brook University, Stony Brook, NY 11794, USA}

\author{Yuxun Guo}
\author{Yoshitaka Hatta  }
\affiliation{Physics Department, Brookhaven National Laboratory, Upton, NY 11973, USA}
\affiliation{RIKEN BNL Research Center, Brookhaven National Laboratory, Upton, NY 11973, USA}

\author{Jakob Schoenleber}
\affiliation{Physik Department T31, James-Franck-Straße 1, Technische Universität München, D-85748 Garching,
Germany}

\begin{abstract}

The familiar Bethe-Heitler process on the proton target $e+p\to e+p+\gamma$ is transformed into a laboratory for studying multiparticle entanglement. We discuss how bipartite and genuine tripartite entanglement between  the final state electron, proton and photon are built up by successive $1\to 2$ and $2\to 2$   elementary interactions. We 
validate our argument by simulating events. Below 5 GeV center-of-mass energy, we identify more than 900 Greenberger-Horne-Zeilinger (GHZ) states and 1200 W states, each  with a fidelty exceeding 99\%.     
\end{abstract}

\maketitle

{\it Introduction}---In 1934, Bethe and Heitler (BH) computed the cross sections for  Bremsstrahlung $e^-+Z\to e^-+\gamma+Z$  and its crossing-symmetric process,  pair creation $\gamma +Z\to e^+ + e^-+Z$, off an atomic nucleus $Z$ \cite{Bethe:1934za,Mo:1968cg}. Their work marks one of the earliest successful applications of Dirac's theory for relativistic fermions. Since then, the two processes have been discussed in many different branches of physics. Pair creation in matter is fundamental to the development of cosmic-ray showers in the atmosphere \cite{Heitler:1937hmt} and   electromagnetic showers in particle  calorimeters \cite{Rossi:1941zza}. The Bremsstrahlung process  finds its modern application in Deeply Virtual Compton Scattering (DVCS) \cite{Ji:1996nm,Diehl:2001pm,Belitsky:2005qn}. In this context,  the photon emission from the incoming or outgoing electron  is often regarded as an unwanted   background that complicates 
 (but sometimes helps)  
the extraction of  Generalized Parton Distributions (GPD). In the present century alone, experimental measurements have been done at a number of facilities such as Jefferson Lab \cite{CLAS:2001wjj}, COMPASS \cite{COMPASS:2018pup}, HERMES \cite{HERMES:2001bob}, HERA \cite{ZEUS:2003pwh,H1:2007vrx}, and in the future at the Electron-Ion Collider \cite{AbdulKhalek:2021gbh}.     

In this paper, for the first time we shed light on the quantum informational  aspects of the BH process,  focusing on the  Bremsstrahlung channel. If the target is the proton $Z=p$, the three particles in the final state, the electron, the proton and the photon all have  two helicity states, and  can be regarded as qubits. 
Depending on the kinematics of scattering and initial polarizations, the three qubits  can be entangled in quantum mechanical sense. Such a study naturally extends the recent works on two-qubit entanglement in electron-proton scattering \cite{Qi:2025onf,Fucilla:2025kit,Cheng:2025zaw,Hatta:2025obw,Fucilla:2026mkg,Agrawal:2026zwa,Bloss:2026yrf,Hatta:2026dqs,Xiao:2026tbs}, and more generally, in high energy scattering in quantum field theory  \cite{Afik:2022kwm,Barr:2024djo,Afik:2025ejh}.  The pattern of entanglement for three-qubit systems is considerably richer and more intricate than for two-qubit systems  \cite{Greenberger:1989tfe,Coffman:1999jd}. Moreover, they play an important role in quantum information science and quantum computing protocols, because they constitute the simplest setting in which genuine multipartite entanglement can arise \cite{Cunha:2019jex,Walter:2016lgl}. However,  tripartite entanglement in elementary particle systems has been less explored 
than bipartite entanglement, and so far mostly in three-body decays of massive (virtual) particles \cite{Acin:2000cs,Hiesmayr:2017xgx,Sakurai:2023nsc,Aguilar-Saavedra:2024whi,Morales:2024jhj,Subba:2024mnl,Fabbrichesi:2025zpw,
Goncalves:2026nnx,Banacki:2026msu}, or certain $2\to 3$ and $3\to 3$ processes \cite{Blasone:2024dud,Sou:2025tyf,Chu:2026yxm,Cao:2026mza} that are not easily accessible in actual experiments. 
Compared to particle decays and $s$-channel processes, where the coherence of the production process naturally leads to entanglement among the three outgoing particles, generating entanglement in $2\to 3$ scattering with $t$-channel exchange is more nontrivial.  We will show that the BH process, one of the most familiar textbook QED processes, offers an interesting laboratory for studying  how multipartite entanglement builds up in successive  interactions of elementary particles. The predictions can, in principle, be tested in experiments equipped with final state spin polarimeters.

{\it Three-qubit entanglement}---It is well established \cite{Dur:2000zz} that  three-qubit entanglement can be classified into four inequivalent classes: 
(i) Fully separable; (ii) Biseparable; (iii) Greenberger–Horne–Zeilinger (GHZ) class \cite{Greenberger:1989tfe}; (iv) W class \cite{Dur:2000zz}. Every pure three-qubit state $ABC$ can be transformed into one of these classes  via stochastic local operations and classical communications (SLOCC) \cite{Dur:2000zz},  consisting of matrix multiplications of the form $M_A\otimes M_B\otimes M_C$ with $M\in$ GL(2,$\mathbb{C}$).    

Fully separable  states   $|\psi_A\rangle \otimes |\psi_B\rangle \otimes |\psi_C\rangle$ are not entangled. We will not be interested in these states in this work. 
Biseparable states contain one entangled pair.  For example, the subsystem $AB$ can be maximally entangled in the form  
\beq
\frac{|+_A-_B\rangle+|-_A+_B\rangle }{\sqrt{2}}\otimes |\psi_C\rangle. \label{12}
\eeq
Such states are characterized by the maximum concurrence \cite{Wootters:1997id} ${\cal C}_{AB}=1$ of the reduced density matrix $\rho_{AB}\equiv {\rm Tr}_{C}\rho$. 
For a pure state, one can also compute   concurrence between one qubit and the other two qubits  as ${\cal C}_{A(BC)}=\sqrt{2(1-{\rm Tr}\rho_A^2)}$,  where $\rho_A={\rm Tr}_{BC}\rho$ \cite{Rungta:2001zcj}. A fundamental property of three-qubit entanglement  is the 
monogamy inequality   $
 {\cal C}_{AB}^2+{\cal C}^2_{AC} \le {\cal C}^2_{A(BC)} \le 1$ \cite{Coffman:1999jd} which tells that, if $A$ is maximally entangled with $B$  (${\cal C}_{AB}=1$), then it cannot be simultaneously entangled with $C$ (${\cal C}_{AC}=0$).

The GHZ state   
\beq
\frac{|+++\rangle+ |---\rangle}{\sqrt{2}}, \label{ghz}
\eeq
has `genuine tripartite entanglement,' and is sometimes described as the  quantum analog of the Borromean ring \cite{Aravind1997}:  Each qubit is entangled with the other two ${\cal C}_{A(BC)}={\cal C}_{B(CA)}={\cal C}_{C(AB)}=1$,  but no two qubits are pairwise entangled ${\cal C}_{AB}={\cal C}_{BC}={\cal C}_{CA}=0$. The GHZ state has the highest degree of genuine tripartite entanglement $F_3^{\rm GHZ}=1$ as measured by the area of the `concurrence triangle' \cite{Jin:2022kxb} 
{\small \beq
 F_3=\!\left[\frac{16}{3}Q(Q-{\cal C}_{A(BC)})(Q-{\cal C}_{B(CA)})(Q-{\cal C}_{C(AB)})\right]^{\frac{1}{2}},\!\! \!\notag
\eeq}with $Q=\frac{1}{2}({\cal C}_{A(BC)}+{\cal C}_{B(CA)}+{\cal  C}_{C(AB)})$.

 The  W state  
\beq
\frac{|+--\rangle+|-+-\rangle+|--+\rangle}{\sqrt{3}} , \label{w}
\eeq
also exhibits a high degree of genuine tripartite entanglement  $F^{\rm W}_3=\frac{8}{9}$, but unlike the GHZ state, entanglement  survives after tracing out one qubit  ${\cal C}_{AB}={\cal C}_{BC}={\cal C}_{CA}=\frac{2}{3}$.  In practice, we will be interested not only in the `canonical' \cite{Acin:2000stw} representations  (\ref{ghz}) and  (\ref{w}), but also their equivalents  $(M_A\otimes M_B\otimes M_C)|\psi_{\rm GHZ/W}\rangle$ with $M \in$ U(2). Local unitary transformations conserve key entanglement measures such as concurrence and $F_3$.

\begin{figure}[t] 
\begin{overpic}[width=0.5\textwidth]{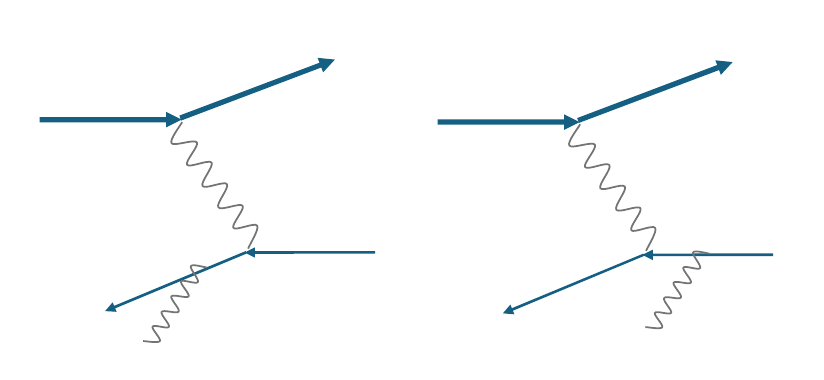}
\put(6,35){$p$}
\put(30,38){$p'$}
\put(10,11){$\ell'$}
\put(41,17){$\ell$}
\put(23,6){$\gamma$}
\put(19,21){$\gamma^*$}
\put(55,35){$p$}
\put(79,38){$p'$}
\put(59,11){$\ell'$}
\put(90,17){$\ell$}
\put(84,8){$\gamma$}
\put(67,21){$\gamma^*$}
\end{overpic}
\vspace{-8mm}
\caption{Feynman diagrams for the Bethe-Heitler (Bremsstrahlung) process}
\label{fig1}
\vspace{-1mm}
\end{figure}
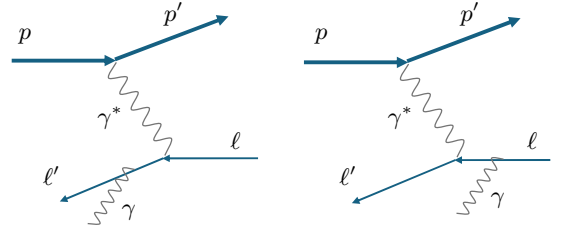

{\it Entanglement in the BH process}---The BH process  $\ell(k,h_0)+p(s_0) \to \ell'(k',h)+p'(s)+\gamma(q,\lambda)$ is defined by the $t$-channel exchange of a virtual photon between the lepton $\ell$ and the proton $p$, and a concomitant emission (Bremsstrahlung) of a real photon $\gamma$ from the initial or final lepton, see Fig.~\ref{fig1}. Emissions from the proton legs are neglected due to the heavy proton mass $m_p$.   We  work in the center-of-mass (CM) frame where the incoming proton and lepton   move in the $\pm z$  directions, respectively.  The final state is a three-qubit system consisting of the scattered lepton $A=\ell'$, the scattered proton $B=p'$ and the produced photon $C=\gamma$. We label these states by their helicities  $|h s \lambda\rangle=|\pm_{\ell'} \pm_{p'} \pm_\gamma\rangle$ (suppressing momentum labels) and  define the helicity amplitudes as 
\beq
&& \hspace{-5mm}\langle hs\lambda|h_0s_0\rangle \equiv \epsilon_\mu^*(q,\lambda)\bar{u}_{\ell'}(k',h) \nn && \hspace{-3mm}\times \Biggl( \frac{2k'^\mu \gamma^\nu+\gamma^\mu\Slash q\gamma^\nu}{2k'\cdot q}  + \frac{2k^\mu\gamma^\nu-\gamma^\nu\Slash q\gamma^\mu}{-2k\cdot q}\Biggr) u_\ell(k,h_0) \nn && \hspace{-3mm}\times \frac{1}{t} \bar{u}(p',s)\left[\gamma_\nu F_1(t)+\frac{i\sigma_{\nu\rho}\Delta^\rho}{2m_p}F_2(t)\right]u(p,s_0), 
\eeq
where $F_{1,2}(t)$ with $t=\Delta^2$, $\Delta^\mu=p'^\mu-p^\mu$ are the electromagnetic form factors of the proton.  
The density matrix of the three-qubit system reads 
\beq
\rho(hs\lambda,h's'\lambda'|h_0s_0)\!=\frac{\langle hs\lambda|h_0s_0\rangle \langle h's'\lambda'|h_0s_0\rangle^*}{\sum_{hs\lambda} \langle hs\lambda|h_0s_0\rangle\langle hs\lambda|h_0s_0\rangle^*}.\notag
\eeq
At fixed initial polarizations $h_0,s_0$, this $8\times 8$ matrix represents a pure final state, from which various concurrences ${\cal C}$ can be calculated.

{\it Bipartite entanglement}---Let us discuss how the different classes of states can be realized in the BH  process.  We begin with bipartite entanglement. 
(i) Lepton-proton maximal entanglement ${\cal C}_{\ell p}\approx 1$   occurs if the initial lepton and proton are both transversely polarized along the same  direction, and when  the scattering is backward (scattering angle $\theta_{p'}\approx 180^\circ$). This was observed in \cite{Cheng:2025zaw} in the context of elastic scattering $\ell p\to \ell'p'$. It can be easily shown that the Bell state 
\beq
|\Phi^\pm\rangle_{\ell' p'}  = \frac{1}{\sqrt{2}}\left(|+_{\ell'}+_{p'}\rangle \pm |-_{\ell'}-_{p'}\rangle \right) ,\label{epbell}
\eeq
is formed in this kinematics, where the plus/minus sign corresponds to the initial polarizations being  parallel/anti-parallel in the transverse plane.  Adding one photon in the final state does not destroy this coherence, as long as the photon  is soft.

(ii) Lepton-photon maximal entanglement ${\cal C}_{\ell \gamma}\approx 1$ occurs when the (timelike) lepton emits a hard photon that carries away a large fraction $z\approx 1$  of the lepton energy, see the first diagram in Fig.~\ref{fig1}. The nonvanishing components of the helicity amplitudes for the splitting $\ell^*\to \ell'+\gamma$ are $\langle+_{\ell'} +_\gamma|+_{\ell^*}\rangle\propto \frac{\sqrt{2(1-z)}}{z(1-z)}$ and $\langle+-|+\rangle\propto \frac{\sqrt{2(1-z)}}{z}$, and their parity partners  \cite{Peskin:1995ev}. Again it is interesting to consider the transversely polarized  incoming lepton. In this case,  the spin correlation matrix  between the final state lepton and photon is calculated as \cite{supple} 
\beq
\langle \sigma^a\otimes \sigma^b\rangle = 
\begin{pmatrix} \pm 1 & 0 & 0 \\ 0 & \frac{\mp (2-z)z}{2-z(2-z)} & 0 \\ 0 & 0 & \frac{(2-z)z}{2-(2-z)z} \end{pmatrix}_{ab}, \label{cma}
\eeq
where the sign depends on the direction of the transverse polarization. In the $z\to 1$ limit the matrix becomes ${\rm diag}(\pm 1, \mp 1, 1)$, representing the Bell state $|\Phi^\pm \rangle_{\ell' \gamma} = \frac{1}{\sqrt{2}}(|+_{\ell'}+_{\gamma}\rangle \pm |--\rangle)$. The proton polarization does not a play a role in this argument. It can be unpolarized, or replaced by a spinless nucleus.  

(iii) Proton-photon entanglement is most  nontrivial because the two particles do not directly interact. To understand this, it is useful to view the BH process as the combination of splitting $p\to p'+\gamma^*$ and (virtual) Compton scattering $\gamma^*+\ell \to \gamma+\ell'$. If the  proton is deflected by a small angle,  the splitting creates the state \cite{supple}
\beq
|\pm_p\rangle \to |\mp_{p'}  \pm_{\gamma^*}\rangle \pm N|\pm_{p'} L_{\gamma^*}\rangle, \label{spl}
\eeq
where $|L_{\gamma^*}\rangle$ denotes the longitudinally polarized photon state with virtuality $t=\Delta^2$.  Although $N$ is typically  large  $|N|\sim m/\sqrt{|t|}> 1$ at small-$|t|$,  the longitudinal Compton scattering amplitude is suppressed $\epsilon_L^\mu {\cal M}^{\gamma^*_L\ell \to \gamma \ell'}_\mu\propto \sqrt{|t|}$. Moreover, it is additionally suppressed  in the forward region $\theta_\gamma \to 0$ due to helicity conservation. It then follows that, if the incoming proton is transversely polarized $|+_p\rangle\pm |-_p\rangle$, the intermediate state is effectively  the Bell state $|\Psi^\pm\rangle_{p\gamma^*}=\frac{1}{\sqrt{2}}(|+_p-_{\gamma^*}\rangle \pm |-+\rangle)$ when  $\theta_\gamma \ll  1$. 
Now assume that the lepton is  longitudinally polarized $\alpha_\ell=0,\frac{\pi}{2}$. The subsequent Compton scattering is dominated by the channels $|\pm_{\ell}+_{\gamma^*}\rangle\to |\pm_{\ell}+_\gamma\rangle$ and $|\pm_{\ell}-_{\gamma^*}\rangle\to |\pm_{\ell}-_\gamma\rangle$     \cite{Hatta:2026dqs},  preserving   $C_{p\gamma}\approx 1$.

{\it Tripartite entanglement}---States with genuine tripartite entanglement can only be generated through two-step processes. For example, first produce a maximally entangled $\ell^* +p$ pair  in the backward scattering region as before (\ref{epbell}).  
Next,  let the lepton emit a hard $z\to 1$ photon. The amplitude $\langle+_\ell+_\gamma|+_\ell^*\rangle$ dominates  over $\langle+_\ell-_\gamma|+_\ell^*\rangle$, and we immediately obtain the GHZ state in its canonical form 
\beq
\!\frac{ |+_{\ell^*}+_{p'}\rangle \pm |--\rangle}{\sqrt{2}}  \to \!\frac{ |+_{\ell'}+_{p'}+_\gamma\rangle \pm |---\rangle}{\sqrt{2}}. \label{ghz1}
\eeq
Another mechanism is to 
combine the splitting (\ref{spl})  with virtual Compton scattering    with deliberately chosen initial polarization vectors. Let us introduce two mixing angles $0\le \alpha_\ell,\alpha_p \le \pi$ 
\beq
|\psi_{\rm in}\rangle_{\ell/p}=\cos\alpha_{\ell/p}|+_{\ell/p}\rangle + \sin\alpha_{\ell/p}|-_{\ell/p}\rangle. \label{alpha}
\eeq
This is not the most general parametrization, as it describes only initial polarizations in the $xz$ plane, but it is sufficient for the present purpose.  
Away from the forward region $\theta_\gamma\gtrsim 1$, the same-helicity scattering $\ket{\pm_\ell \pm_{\gamma^*}}\to\ket{\pm_{\ell'} \pm_{\gamma}}$ dominates over the opposite-helicity scattering $\ket{\pm_\ell \mp_{\gamma^*}}\to\ket{\pm_{\ell'} \mp_\gamma}$ \cite{Hatta:2026dqs}.   
 In the very backward region $\theta_\gamma \to \pi$ where  $k\cdot q\sim m_\ell^2$, the lepton  helicity-flip scattering $\ket{+_\ell-_{\gamma^*}}\to\ket{-_{\ell'}+_\gamma}$ dominates. Thus the final state is approximately 
\beq
\label{ghz2}
\cos\alpha_p\cos\alpha_\ell\ket{+_{\ell'}-_{p'}+_\gamma}
+\sin\alpha_p\sin\alpha_\ell\ket{-+-}\, , 
\eeq
 when $\theta_{\gamma}\gtrsim 1$   and 
\beq
\label{ghz3}
\cos\alpha_p\sin\alpha_\ell\ket{+_{\ell'}-_{p'}-_\gamma}
+\sin\alpha_p\cos\alpha_\ell\ket{-++},
\eeq
when $\theta_\gamma \approx \pi$. 
A GHZ state is formed when the magnitudes of the two components are equal.  
This implies $|\tan\alpha_p|\simeq|\tan\alpha_\ell|^{-1}$ ($\alpha_p=\pm \alpha_\ell\pm(2n+1) \frac{\pi}{2}$) and  $|\tan\alpha_p|\simeq|\tan\alpha_\ell|$ ($\alpha_p=\pm \alpha_\ell+n\pi$)  in the two cases.

Finally, the  W state (\ref{w}) occurs when the longitudinal photon component $|L_{\gamma^*}\rangle$ in (\ref{spl}) is not negligible. 
Assume that the proton has plus helicity $|+_p\rangle$ $(\alpha_p=0$) and the lepton is transversely polarized ($\alpha_\ell=\frac{\pi}{4}$). The initial condition for the virtual Compton scattering is then 
\beq
 (|+_\ell\rangle+|-_\ell\rangle) \otimes \left( |-_{p'}+_{\gamma^*}\rangle+N|+_{p'}L_{\gamma^*}\rangle\right) .
\eeq
When $\theta_\gamma\gtrsim 1$, $|+_{\gamma^*}\rangle$ dominantly scatters with  $|+_\ell\rangle$, while $|L_{\gamma^*}\rangle$ scatters with $|\pm_\ell\rangle$ with equal strength, producing  $|\pm_{\ell'} \mp_{\gamma^*} \rangle$. 
Then the final state takes the form 
\beq
N'|+_{\ell'}-_{p'}+_\gamma\rangle +  N\left(|++-\rangle+ |-++\rangle \right). \label{he}
\eeq 
If it is possible to tune the   scattering parameters such that $|N'|=|N|$.  (\ref{he}) becomes the W state up to a local unitary transformation.

\begin{figure}[t]
    \centering
\includegraphics[width=0.8\linewidth]{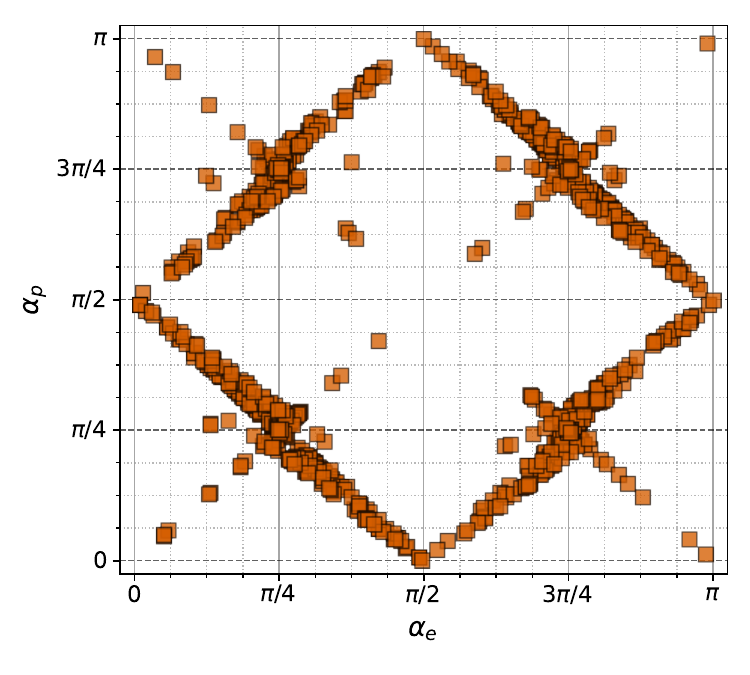 } \\ \includegraphics[width=0.8\linewidth]{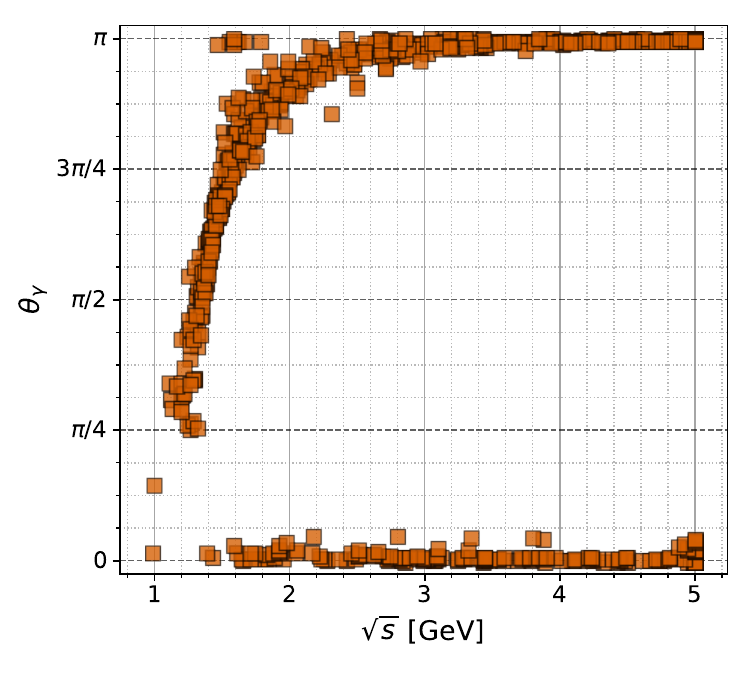}
    \caption{GHZ states in the $(\alpha_e,\alpha_p)$ plane (top) and $(\sqrt{s},\theta_\gamma)$ plane (bottom) }
    \label{ghzfig}
\end{figure}

\begin{figure}[t]
    \centering
    \includegraphics[width=0.8\linewidth]{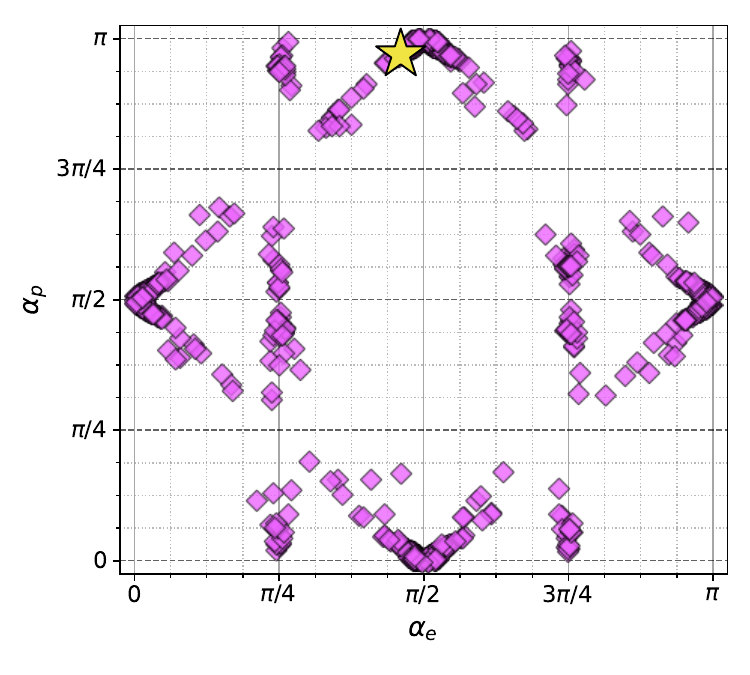} \\ \includegraphics[width=0.8\linewidth]{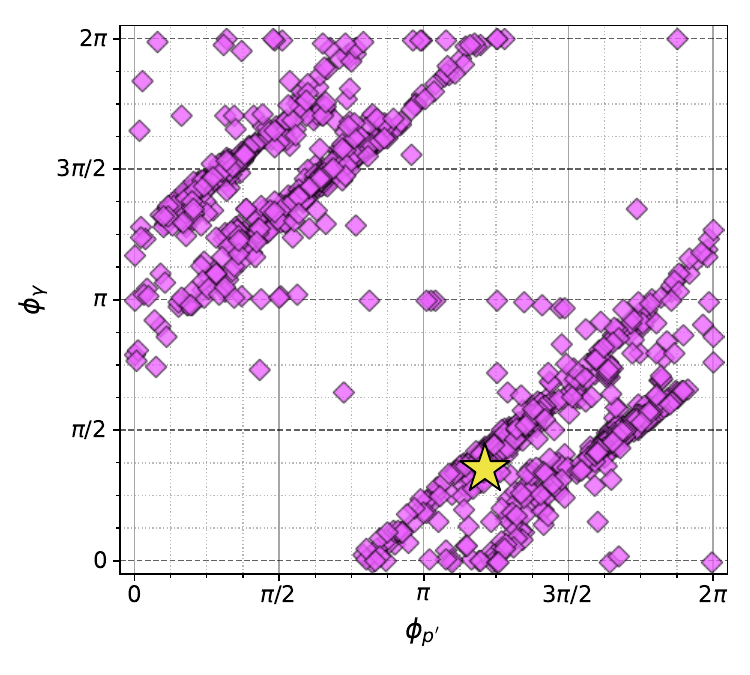}
    \caption{W states in the $(\alpha_e,\alpha_p)$ plane (top) and $(\phi_{p'},\phi_\gamma)$ plane (bottom)}
    \label{wfig}
\end{figure}

{\it Simulations}---To validate the qualitative arguments presented above, and to find states of more complicated origin, we now perform numerical simulations. For definiteness, we consider an electron beam $\ell =e$. The proton form factors $F_{1,2}(t)$ are taken from \cite{Ye:2017gyb}. The independent kinematic variables are  the CM energy $\sqrt{s}$, the polar angles of the outgoing proton ($\theta_{p'},\phi_{p'})$ and the photon ($\theta_{\gamma},\phi_{\gamma})$, the  photon energy $E_\gamma$, and the two mixing angles $\alpha_{e/p}$ introduced in (\ref{alpha}). 

To identify states locally equivalent to the canonical GHZ and W states, (\ref{ghz}) and (\ref{w}), we minimize the following invariant loss functions
\beq
D^2_{\rm GHZ} =
{\cal C}_{e p}^2
+{\cal C}_{p\gamma}^2+{\cal C}_{e\gamma}^2
+\left(F_3-1\right)^2,
\eeq
for the GHZ class and 
\beq
D^2_{\rm W} \equiv
\left({\cal C}_{e p}-\frac{2}{3}\right)^2\!
+\left({\cal C}_{p\gamma}-\frac{2}{3}\right)^2\!
+\left({\cal C}_{e\gamma}-\frac{2}{3}\right)^2\!,
\eeq
for the W class. The losses are searched over the eight-dimensional parameter space. We optimized from 2,048 different starting guesses: $1024$ evenly distributed guesses~\cite{mckay1979} together with $1024$ guesses selected from $32768$ scrambled Sobol
points~\cite{SOBOL196786}. Each starting point is refined using the bounded L-BFGS-B algorithm~\cite{Zhu:1997plu} implemented in the \texttt{SciPy} package~\cite{2020SciPy-NMeth}, followed by a multiscale direct-search step that verifies the local minimum and accounts for periodic angular boundaries. Solutions converging to the same basin are deduplicated. We retain candidates satisfying $D_{\rm{GHZ/W}}<0.01$. Finally, each candidate state $|\psi\rangle$ is independently verified through its local-unitary fidelity,
\beq
F=\max\limits_{M_{e,p,\gamma}}\left|\langle {\rm GHZ/W}|(M_e\otimes M_p\otimes M_\gamma)|\psi\rangle\right|^2.
\eeq
The three ${\rm SU}(2)$ transformations $M_{e,p,\gamma}$ are parameterized by nine Euler angles and optimized using twelve L-BFGS-B restarts. Candidates with $F_{\rm{GHZ/W}}>0.99$ pass the selection.

The results for $\sqrt{s}\le5$ GeV are shown in Fig.~\ref{ghzfig} and Fig.~\ref{wfig}, where each dot represents a GHZ or a  W state (local minimum), respectively.   These are two-dimensional projections of the eight-dimensional phase space, so `nearby' dots are not necessarily close in the full parameter space.  (More plots and details are given in \cite{supple}.) In total, we have identified 949 GHZ minima and 1255 W minima well-isolated in phase space. Many of them appear in the kinematical regions predicted by our arguments. In particular,  GHZ states with $\theta_\gamma\approx 0$, $\theta_\gamma\sim 1$ and $\theta_\gamma\approx \pi$ roughly correspond to (\ref{ghz1}), (\ref{ghz2}) and  [The states of the form (\ref{ghz2}) are much rarer because one requires fine-tuning the parameters to suppress the longitudinal photon contribution when $\theta_\gamma \sim 1$.]  Also the distribution of the mixing angles $\alpha_{e,p}$ is consistent with (\ref{ghz2}) and (\ref{ghz3}).  However,  many other states do not admit a simple explanation. Most of these unexpected states are in `non-canonical' forms. For example, the  W state marked by a star in Fig.~\ref{wfig} occurs at
$\sqrt{s}\approx1.48~\mathrm{GeV}$, $E_\gamma\approx0.213~\mathrm{GeV}$,
$\theta_{p'}\approx0.711\pi$, $\theta_\gamma\approx0.592\pi$,
$\phi_{p'}\approx1.21\pi$, $\phi_\gamma\approx0.351\pi$,
$\alpha_e\approx0.460\pi$, and $\alpha_p\approx0.970\pi$. 
Its normalized wavefunction reads 
{\small \beq
\begin{aligned}
& 0.543e^{-0.260\pi\mathrm{i}}|+_{e'}+_{p'}+_{\gamma}\rangle +0.259e^{-0.107\pi\mathrm{i}}|++-\rangle \\ 
& +0.0273e^{0.925\pi\mathrm{i}}|+-+\rangle +0.0732e^{0.908\pi\mathrm{i}}|+--\rangle 
\\
&+0.252e^{-0.660\pi\mathrm{i}}|-++\rangle+0.459e^{0.266\pi\mathrm{i}}|-+-\rangle
\\
&+0.505e^{-0.532\pi\mathrm{i}}|--+\rangle 
+ 0.321e^{-0.817\pi\mathrm{i}}|---\rangle.
\end{aligned} \label{example}
\eeq}Via the following SU(2) matrices 
\beq
M_e&=
\begin{pmatrix}
0.948e^{0.00475\pi i} &
0.318e^{0.972\pi i}\\
-0.318e^{-0.972\pi i} &
0.948e^{-0.00475\pi i}
\end{pmatrix},\nonumber
\\
M_p&=
\begin{pmatrix}
0.310e^{0.385\pi i} &
0.951e^{0.690\pi i}\\
-0.951e^{-0.690\pi i} &
0.310e^{-0.385\pi i}
\end{pmatrix}, \nonumber
\\ 
M_\gamma&=
\begin{pmatrix}
0.395e^{-0.830\pi i} &
0.919e^{0.269\pi i}\\
-0.919e^{-0.269\pi i} &
0.395e^{0.830\pi i}
\end{pmatrix},
\eeq
(\ref{example}) can be rotated to the canonical form (\ref{w}) up to an overall phase with fidelity $F_{\rm W}>0.999$.

In \cite{supple}, we show the  distribution of states with maximal bipartite entanglement $1-{\cal C}_{ij} \approx 0$ ($ij=ep,e\gamma,p\gamma$). Such states are far more densely populated in phase space than tripartite entangled states.  
 Around each local minimum, we define `correlation lengths' in the eight dimensional phase space   over which $D_{\rm GHZ/W}$ or $1-{\cal C}_{ij}$ is degraded by 0.02. For GHZ and W states, the correlation  lengths in the momentum directions are typically ${\cal O}(10)$ MeV, as compared to ${\cal O}(100)$ MeV for bipartite entangled states \cite{supple}. 
Consequently, the GHZ and W states occupy phase space volumes roughly $10^{-8}$ times smaller than that of the most abundant ${\cal C}_{p\gamma}\approx 1$ states.

{\it Conclusions}---We have presented the first study of quantum entanglement in the Bethe-Heitler process. This important but mundane process exhibits a surprisingly rich pattern of entanglement, including more than 900 GHZ states and 1200 W states, each with a fidelty better than 99\%. Since the first realization of the GHZ  state in a three-photon system  \cite{Bouwmeester:1998iz}, preparing such highly entangled tripartite systems in controlled environments has become routine in quantum information experiments. However, their realizations in elementary particle scattering are far more unusual. Previously, a pair of GHZ states have been found in rare Higgs decay modes $H\to \ell^+\ell^-V$ in the collinear kinematics \cite{Morales:2024jhj,Banacki:2026msu}, whereas not a single W state has been reported to date.   

At higher CM energies,  or in specific corners of phase space such as $(k-k')^2\to 0$, corrections from DVCS \cite{Diehl:2001pm,Belitsky:2005qn}, or more generally,  the reaction $p+\gamma^*\to p'+\gamma$    may need to be included. This will mostly affect the regions ${\cal C}_{p\gamma}\approx 1$, making photon-proton  entanglement  sensitive to the partonic structure of the proton  \cite{Hatta:2026dqs}. It is also interesting to extend the present analysis to spin-$\frac{1}{2}$ nuclei such as $\,^3$He \cite{EPIOSScientificConsortium:2025dfc,Xiao:2026tbs}. We leave these problems to future work. 

From an experimental perspective, it is advantageous   that many interesting states are found in the collinear regions where the cross section is  enhanced. 
Other states are populated at extreme kinematics (such as $\theta_{p'}\approx \pi$), and are more difficult to search. Spin correlations in the final state are in principle measurable via detectors serving as polarimeters. Recently there have been active discussions on final-state photon and proton polarimetry \cite{Lv:2026qmu,Cheng:2026ktp,BessidskaiaBylund:2022qgg,BESIII:2026yyv}. However, measuring the polarization of final-state electrons is more challenging.  Switching to a polarized muon beam, as has traditionally been done at CERN \cite{EuropeanMuon:1987isl,COMPASS:2018pup} (see also \cite{Acosta:2021qpx}), appears to be a promising way forward.  

{\it Acknowledgments}---Y.~G. and Y.~H. were supported by the U.S. Department
of Energy under Contract No. DE-SC0012704, and also by LDRD funds from Brookhaven Science Associates. H.~C. is supported by the U.S. Department of Energy, Office of High Energy Physics, under contract No.~DE-SC0010143, and is partially supported by a CFNS Joint Postdoctoral Fellowship. H.~C. also acknowledges the support in part through the computational resources and staff contributions provided for the Quest high performance computing facility at Northwestern University which is jointly supported by the Office of the Provost, the Office for Research, and Northwestern University Information Technology. J.S. was supported by the Excellence Cluster ORIGINS funded by the Deutsche Forschungsgemeinschaft under
Grant No.~EXC - 2094 - 390783311. Claude and OpenAI Codex were used to assist with numerical code development, with AI-assisted code independently reviewed, tested, and validated by the authors.

\bibliography{ref}

\pagebreak
\widetext
\begin{center}
\textbf{\large Supplemental Materials}
\end{center}
\setcounter{section}{0}
\setcounter{equation}{0}
\setcounter{figure}{0}
\setcounter{table}{0}
\setcounter{page}{1}
\makeatletter
\renewcommand{\theequation}{S\arabic{equation}}
\renewcommand{\thefigure}{S\arabic{figure}}

%

\onecolumngrid

\section{Helicity amplitudes in $p\to p'+\gamma^*$}

\begin{figure}[h]
    \centering
    \includegraphics[width=0.4\linewidth]{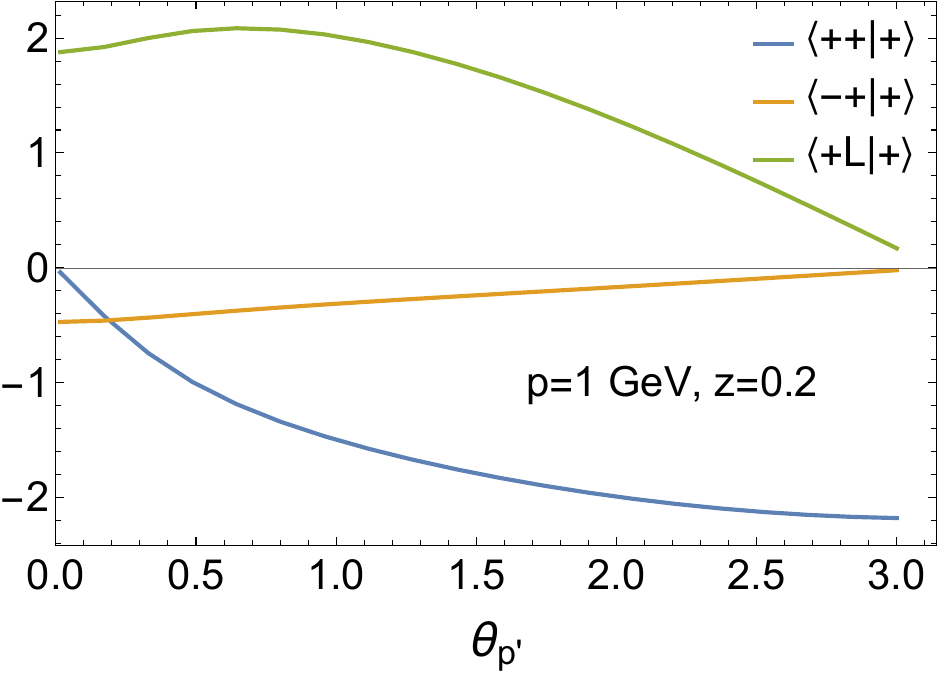}
    \caption{Helicity amplitudes in $p\to p'+\gamma^*$  as a function of the proton deflection angle $\theta_{p'}$.}
    \label{ppgamma}
\end{figure}

Here we study the helicity structure of the splitting  $p\to p'+\gamma^*$. We parametrize the initial and final momenta as     
\beq
&&p^\mu=(E,0,0,p), \qquad p'^\mu=((1-z)E,p'\sin\theta_{p'},0,p'\cos\theta_{p'}), \nn
&& \qquad \Delta^\mu=p'^\mu - p^\mu=(zE,-p'\sin\theta_{p'},0,p-p'\cos\theta_{p'}),
\eeq
where $p'=\sqrt{(1-z)^2E^2-m^2}$. The  photon has virtuality  
\beq
-t=-\Delta^2
= 2(1-z)p^2-2zm^2-2pp'\cos\theta_{p'},
\eeq
and three polarization vectors 
\beq
\epsilon^{*\mu}_\pm = \frac{1}{\sqrt{2}}\left(0, \frac{p-p'\cos\theta_{p'}}{|\vec{\Delta}|},\mp i, \frac{p'\sin\theta_{p'}}{|\vec{\Delta}|}\right), \qquad \epsilon^\mu_L=\frac{1}{\sqrt{-t}}\left( |\vec{\Delta}|, \frac{ -p'\sin\theta_{p'} }{|\vec{\Delta}|}zE,0, \frac{p-p'\cos\theta_{p'}}{|\vec{\Delta}|}zE\right),
\eeq
where $|\vec{\Delta}|=\sqrt{p^2+p'^2-2pp'\cos\theta_{p'}}$ and $L$ denotes `longitudinal.' 
It is straightforward to evaluate the emission amplitudes   in helicity space ($\lambda=\pm,L$) 
\beq
\langle s\lambda|s_0\rangle=\epsilon_{\lambda}^{*\mu} \bar{u}(p',s)\gamma_\mu u(p,s_0)
\eeq
In Fig.~\ref{ppgamma}, we plot $\langle +_{p'}+_{\gamma^*}|+_p\rangle$, $\langle -+|+\rangle$ and $\langle +L|+\rangle$ evaluated at $p=1$ GeV and $z=0.2$. In the forward region $\theta_{p'}\approx 0$, the production of a longitudinally polarized photon is dominant. One can actually show that, at $\theta_{p'}=0$, $\langle +L|+\rangle = -\frac{\sqrt{2}m}{\sqrt{-t}}\langle -+|+\rangle$. In the backward region $\theta_{p'}\approx \pi$, the state $|+_{p'}+_{\gamma^*}\rangle$ dominates.

As for the splitting $\ell\to \ell'+\gamma$,  the spin correlation matrix between $\ell'$ and $\gamma$ for a transversely polarized initial lepton is calculated as 
\beq
C_{ab}= \frac{{\rm Tr}[ (\sigma^a \otimes \sigma^b)T \left(\frac{\mathbb{I}\pm \sigma^x}{2}\right) T^\dagger ]}{{\rm Tr}[T \left(\frac{\mathbb{I}\pm \sigma^x}{2}\right) T^\dagger ]} = 
\begin{pmatrix} \pm 1 & 0 & 0 \\ 0 & \frac{\mp (2-z)z}{2-z(2-z)} & 0 \\ 0 & 0 & \frac{(2-z)z}{2-(2-z)z} \end{pmatrix}_{ab},
\eeq
where 
\beq
T= \begin{pmatrix} \langle ++|+\rangle & 0 \\ \langle +-|+\rangle & 0 \\ 0 & \langle -+|-\rangle \\ 0 & \langle --|-\rangle
\end{pmatrix}. 
\eeq
In deriving this, we have set $m=0$ and expanded to linear order in $\theta_{e'}$.

\section{Regions of bipartite and tripartite entanglements below $\sqrt{s}=5$ GeV}

\begin{figure}
    \centering
\includegraphics[width=0.98\linewidth]{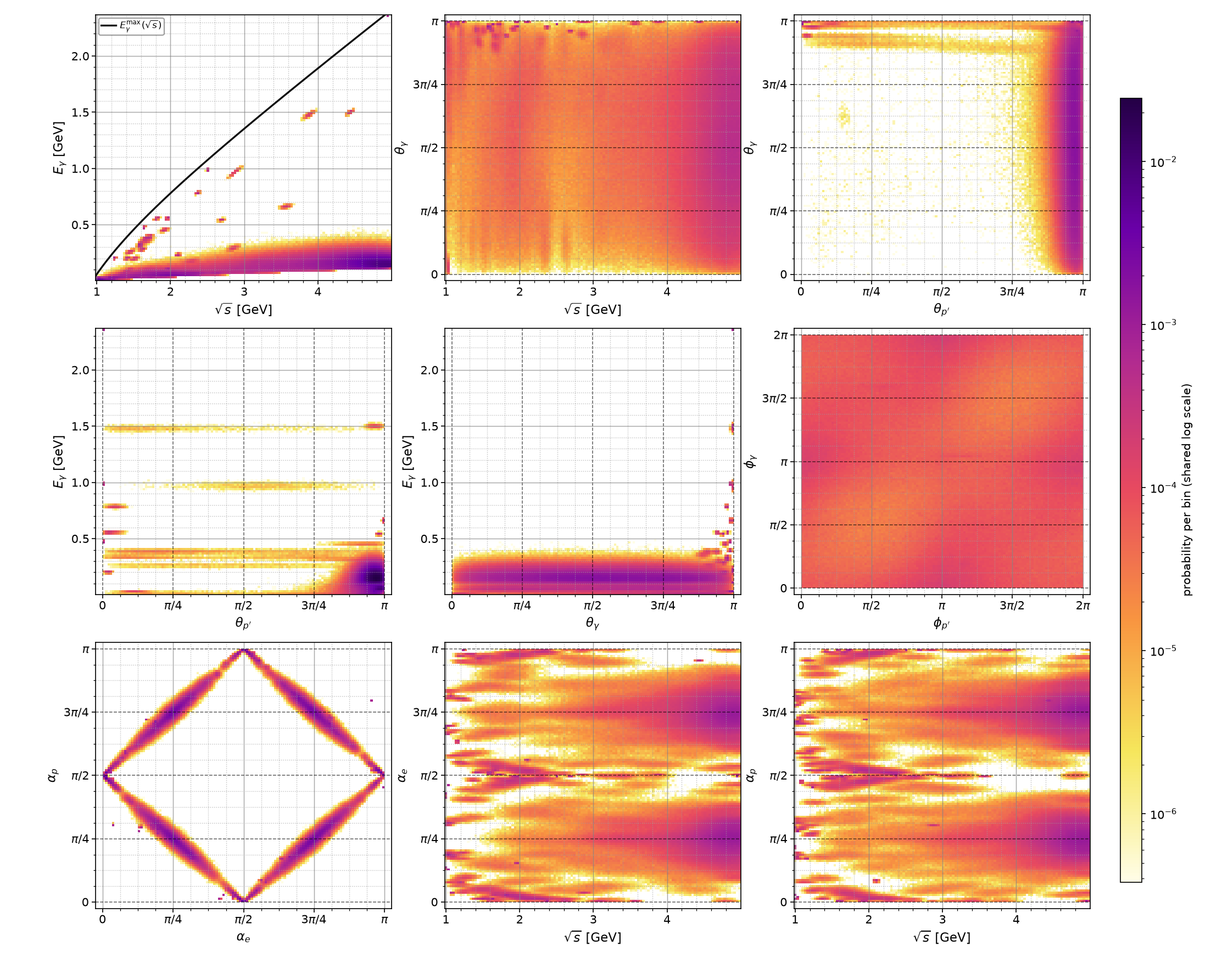}
    \caption{States with ${\cal C}_{ep}\gtrsim 0.97$. Generated from 1,311 identified local minima.}
    \label{ep}
\end{figure}

\begin{figure}
    \centering
\includegraphics[width=0.98\linewidth]{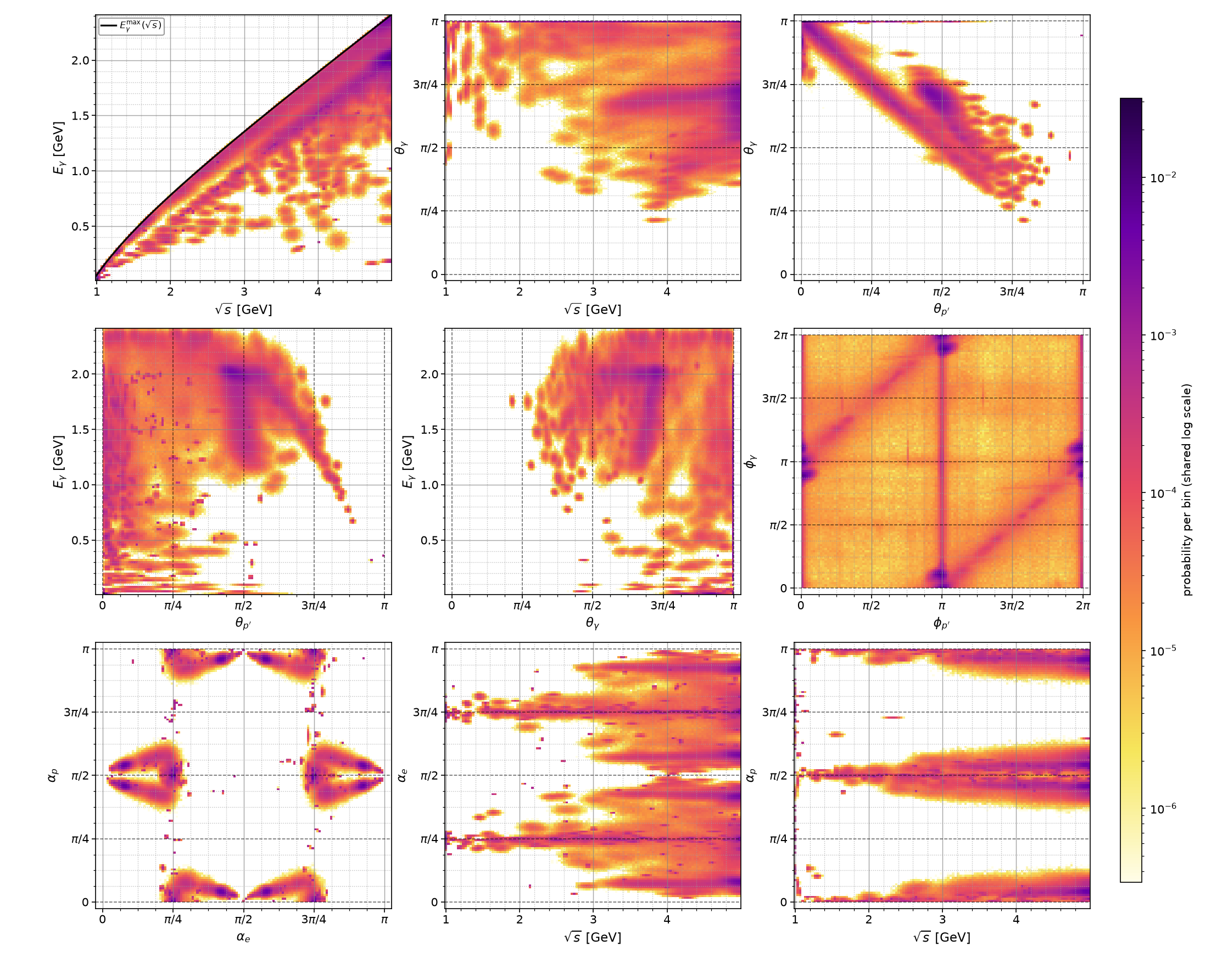}
    \caption{States with ${\cal C}_{e\gamma}\gtrsim 0.97$. Generated from 1,465 identified local minima.}
    \label{egamma}
\end{figure}

\begin{figure}
    \centering
\includegraphics[width=0.98\linewidth]{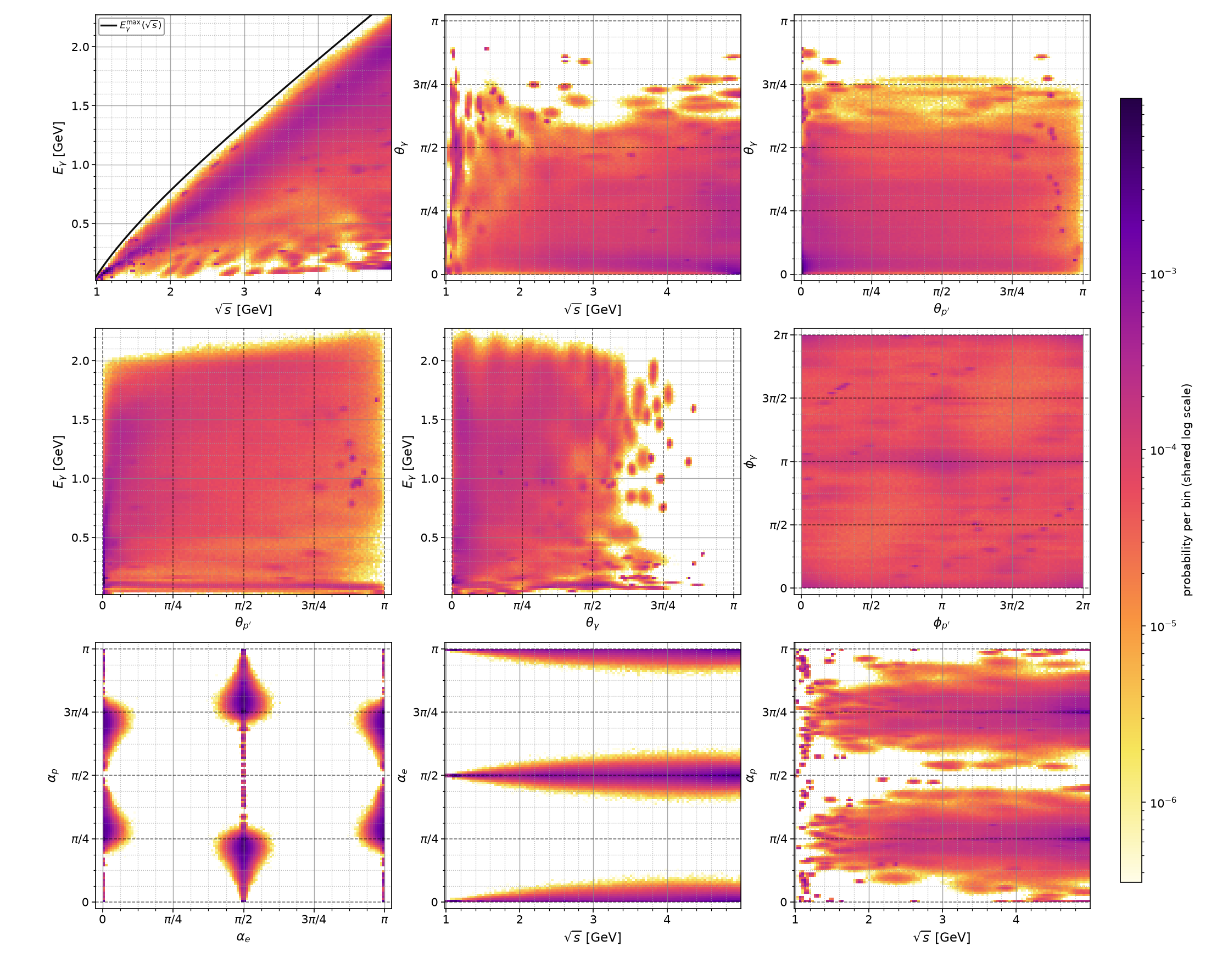}
    \caption{States with ${\cal C}_{p\gamma}\gtrsim 0.97$. Generated from 1,415 identified local minima.}
    \label{pgamma}
\end{figure}

\begin{figure}
    \centering
    \includegraphics[width=0.98\linewidth]{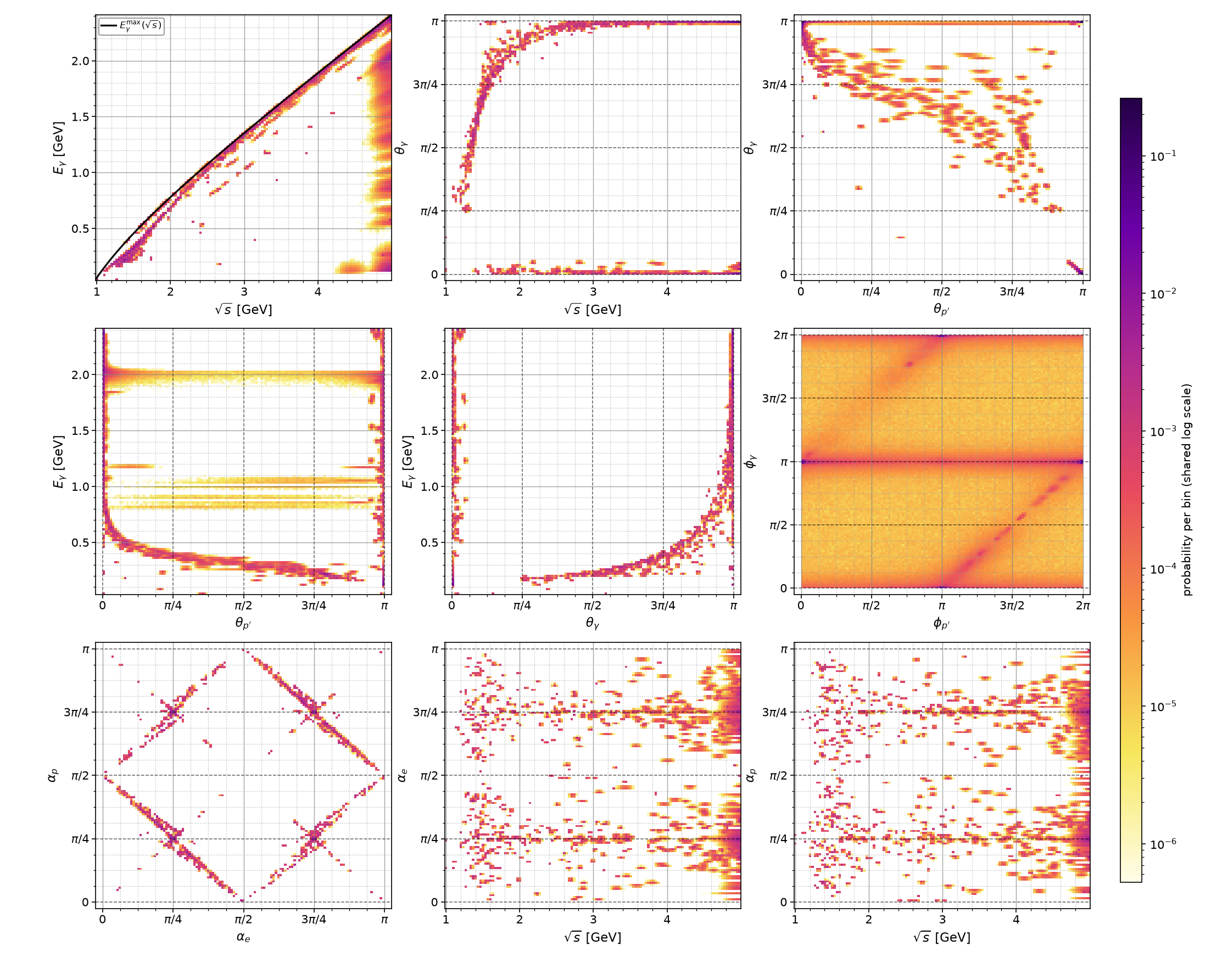}
    \caption{Summary of the GHZ states with 949 identified local minima.}
    \label{fig:GHZsum}
\end{figure}

\begin{figure}
    \centering
    \includegraphics[width=0.98\linewidth]{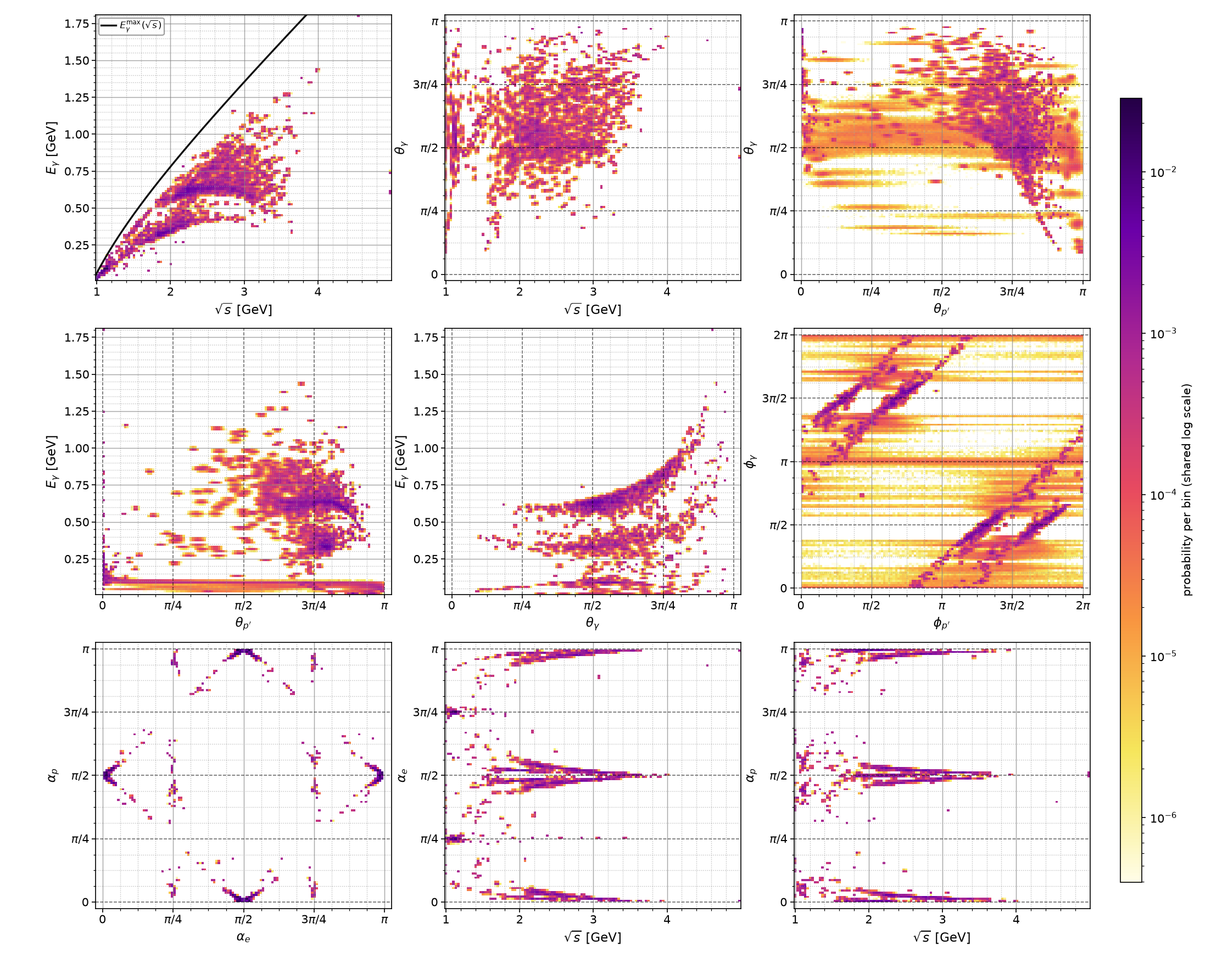}
    \caption{Summary of the W states with 1255 identified local minima.}
    \label{fig:Wsum}
\end{figure}

Here we report all identified regions with strong bipartite and tripartite entanglement below $\sqrt{s}=5$ GeV.  Figs.~\ref{ep},~\ref{egamma},~\ref{pgamma} present Gaussian-smeared histograms for high bipartite concurrences $\mathcal{C}_{ep}$, $\mathcal{C}_{e\gamma}$ and $\mathcal{C}_{p\gamma}$, respectively. Figs.~\ref{fig:GHZsum}  and~\ref{fig:Wsum} are the corresponding plots  for GHZ and W states.

Candidate configurations are first identified using the same local-minimum search procedure described in the main text, retaining only minima satisfying $1-\mathcal{C}_{ij}<0.01$ or $D_{\rm GHZ/W}<0.01$. Minima whose associated phase-space regions overlap by more than $75\%$ are treated as duplicates, with only one representative retained. Around each retained minimum, $2000$ points are then sampled from a multivariate Gaussian distribution, whose covariance matrix is determined numerically using a characteristic width defined by $\Delta\mathcal{C}$ or $\Delta D_{\rm GHZ/W}=0.02$. Including the minima themselves, the resulting ensemble contains $N_{\mathrm{min}}\times(2000+1)$ points, typically satisfying $1-\mathcal{C}_{ij}<0.03$ or $D_{\rm GHZ/W}<0.03$. Histograms are constructed by dividing the range of each parameter into 120 bins, with the normalized density in each bin given by the number of sampled points in that bin divided by the total number of points. 

These histograms should not be interpreted as rigorous or unbiased phase-space probability densities, since they are constructed from a finite sample centered on local minima that depends on the optimization processes and number of samples. Nevertheless, the Gaussian smearing provides a straightforward representation of the local phase-space structure and makes correlations among parameters more easily identifiable than in the discrete set of minima alone.

In Fig.~\ref{ep}, the region of maximal ${\cal C}_{ep}\approx 1$ is consistent with our expectation  $\theta_{p'}\approx \pi$,  $E_\gamma \approx 0$. We have assumed that the proton and the electron are both transversely polarized ($\alpha_{e,p}=\frac{\pi}{4},\frac{3\pi}{4}$), but this condition  can be generalized to  $\alpha_e=\pm \alpha_p\pm (2n+1)\frac{\pi}{2}$. 

Fig.~\ref{egamma} also confirms our expectation. ${\cal C}_{e\gamma}$ is enhanced when the electron is transversely polarized ($\alpha_e=\frac{\pi}{4},\frac{3\pi}{4}$), the outgoing electron and the photon are collinear $\phi_\gamma =\phi_{p'}\pm \pi$, and the photon is energetic $E_\gamma \sim \sqrt{s}/2$. However, there are other regions where ${\cal C}_{e\gamma}$ is enhanced. In particular, it was unexpected that longitudinal proton polarization $\alpha_p=0,\frac{\pi}{2}$ is preferred.


Fig.~\ref{pgamma}  bears out our argument that  ${\cal C}_{p\gamma}$  is enhanced when the proton is transversely polarized $\alpha_p = \frac{\pi}{4},\frac{3\pi}{4}$ and the electron is  longitudinally polarized $\alpha_e=0,\frac{\pi}{2}$. Besides, the scattering angle $\theta_\gamma$ of the Compton subprocess $\gamma^*+e\to \gamma+e'$  cannot be too large. Contrary to our expectation, however, minima with ${\cal C}_{p\gamma}\approx 1$ are much more frequently encountered than those with ${\cal C}_{ep}\approx 1$ and ${\cal C}_{e\gamma}\approx 1$.  


In Figs.~\ref{fig:GHZsum} and \ref{fig:Wsum}, we show more plots of GHZ and W states.  We find that  W states tend to be more   populated in the small-$\sqrt{s}$ region, while the distribution of GHZ states is roughly independent of $\sqrt{s}$.  In total, we have identified 949 GHZ minima and 1255 W minima. 
One representative example of a GHZ state is located at $\sqrt{s}\approx4.97~\mathrm{GeV}$, $E_\gamma\approx2.17~\mathrm{GeV}$,
$\theta_{p'}\approx0$, $\theta_\gamma\approx\pi$,
$\alpha_e\approx0.241\pi$, and $\alpha_p\approx0.759\pi$. 
Its wavefunction takes the form
\beq
\left|\widetilde{\psi}_{\rm GHZ}\right\rangle
\simeq
-
0.707e^{0.192\pi i}\,|+_{e'}-_{p'}-_{\gamma}\rangle+0.707e^{-0.192\pi i}\,|-_{e'}+_{p'}+_{\gamma}\rangle
. \label{exghz}
\eeq
Via the following local unitary transformations
\beq
M_{\ell'}=
\begin{pmatrix}
0 & 
-e^{-i \beta_\ell}\\e^{i\beta_\ell } 
& 0
\end{pmatrix}, \qquad 
M_{p'}=
\begin{pmatrix}
e^{i\beta_p } & 0\\
0 & e^{-i \beta_p}
\end{pmatrix}, \qquad 
M_\gamma=
\begin{pmatrix}
e^{ i\beta_\gamma} & 0\\
0 & e^{-i \beta_\gamma}
\end{pmatrix},
\eeq
with $\beta_\gamma+\beta_p=\beta_l+0.192\pi$, this state is  rotated to the canonical form (2) up to an overall phase with fidelity $F_{\rm GHZ}>0.999$. In contrast to W states (see (17)), GHZ states typically show up  in the two-term form such as (\ref{exghz}). 

Comments are in order regarding the symmetries of the $(\alpha_e,\alpha_p)$ plane (see the lower-left panel in Figs.~\ref{ep}-\ref{fig:Wsum}).  The symmetry under $(\alpha_e,\alpha_p)\to (\pi-\alpha_e,\pi-\alpha_p)$ can be understood as a consequence of parity. The symmetry under $(\alpha_e,\alpha_p)\to (\pi-\alpha_e,\alpha_p)\to (\alpha_e,\pi-\alpha_p)$ is approximate, and we attribute it  to the chiral symmetry of the electron. For a massless electron, its helicity is conserved during scattering. Thus the states $|+_e\rangle$ and $|-_e\rangle$ scatter independently, and their relative phase can be rotated away by a local unitary transformation. Taking  both symmetries into account,  one could reduce the effective range of parameters to $0\le \alpha_{e,p}\le \frac{\pi}{2}$. Nevertheless, our search algorithm is unaware of these symmetries, and actually, almost all the GHZ/W states remain independent even after considering the symmetry transformations   (949$\to$943 for GHZ  states and 1255$\to$ 1255 for W states).

It is obvious by inspection that the phase space density of tripartite entangled states is much lower than that of bipartite entangled states. To make this observation more quantitative, in Table I, we summarize the `correlation lengths' defined as the average ranges of parameters ($p'_{x,y,z}$, $q^{\gamma}_{x,y,z}$, $\alpha_{e,p}$)  over which $1-\mathcal{C}_{ij}$ or $D_{\rm{GHZ/W}}$ is degraded by $\Delta\mathcal{C}$ or $\Delta D_{\rm GHZ/W}=0.02$.  
(Here we switch to Cartesian coordinates because the correlation lengths cannot be meaningfully defined for azimuthal angles $\phi_{p',\gamma}$ at degenerate points $\theta_{p',\gamma}=0,\pi$.) This shows that rather severe fine-tunings are required to detect GHZ/W states. For example, to find a GHZ state, typically momenta have to be measured within an  accuracy of ${\cal O}(10)$ MeV, $\alpha_{p}$ within ${\cal O}(10^{-2})$ radians, etc. The total phase space volume of each class of states is obtained by multiplying the eight correlation lengths with Lorentz invariant volume $(2E_{e'} 2E_{p'}2E_{\gamma})^{-1}$ at each minimum $i$ and summing over all minima $V_{\Sigma}=\sum_i V_i$ in the same class. 
From the union volume $V_{\rm{tot}}$ that removes from $V_\Sigma$  overlaps between  different minima, we roughly estimate that the total phase space volumes of the GHZ and W states are about $10^{-8}$ times smaller than that of the ${\cal C}_{p\gamma}$ states.  We also observe that the overlap rate  $P\equiv 1-V_{\rm{tot}}/V_{\Sigma}$ is large for ${\cal C}_{p\gamma}$ and ${\cal C}_{ep}$, and noticeable  for ${\cal C}_{e\gamma}$, see Table I. This implies that the identified regions of bipartite entanglement form a continuous patch of phase space, with a characteristic extent set by the large correlation lengths.  Adding  more minima does not open up new,  non-overlapping regions of phase space. 
The tripartite entanglement regions, on the contrary, have much smaller correlation lengths and overlap rates. Therefore, each minimum  represents a well-isolated GHZ/W state.

\begin{table*}[b]
\centering
\caption{Values of the physical correlation lengths averaged over the minima of each entanglement measurement and for each Cartesian final-momentum coordinate. Momentum components are in GeV; $\alpha_e,\alpha_p$ are in radian. The overlap rate $P\equiv 1-V_{\rm{tot}}/V_{\Sigma}$ measures how much overlap all the regions have on average.
\label{tab:physical_correlation_lengths}}
\scriptsize
\setlength{\tabcolsep}{3.2pt}
\renewcommand{\arraystretch}{1.15}
\begin{ruledtabular}
\begin{tabular}{lccccccccc}
\textbf{State}
& $p'_{x}$ [GeV]
& $p'_{y}$ [GeV]
& $p'_{z}$ [GeV]
& $q^\gamma_{x}$ [GeV]
& $q^\gamma_{y}$ [GeV]
& $q^\gamma_{z}$ [GeV]
& $\alpha_{e}$
& $\alpha_{p}$
& $P$
\\
\hline
W           & 0.017654 & 0.017372 & 0.019536 & 0.010738
               & 0.010475 & 0.010769 & 0.005126 & 0.006104 & 0.7\% \\
$\mathrm{GHZ}$ & 0.012364 & 0.012401 & 0.037670 & 0.008590
               & 0.008618 & 0.028182 & 0.007246 & 0.007201 & 0.1\% \\
${\cal C}_{ep}$       & 0.377186 & 0.379379 & 0.386784 & 0.165737
               & 0.168909 & 0.160470 & 0.105288 & 0.105227 & 64.1\% \\
${\cal C}_{e\gamma}$  & 0.196942 & 0.208602 & 0.205545 & 0.110170
               & 0.118616 & 0.122175 & 0.065484 & 0.065652 & 14.7\% \\
${\cal C}_{p\gamma}$  & 0.370469 & 0.370095 & 0.427345 & 0.251368
               & 0.252273 & 0.240150 & 0.145201 & 0.122467 & 54.4\% \\
\end{tabular}
\end{ruledtabular}
\end{table*}

\end{document}